\documentclass[twoside,fleqn]{article}
\usepackage{espcrc2}
\usepackage{graphicx}
\newcommand{\be}{\begin{equation}}
\newcommand{\ee}{\end{equation}}

\title{
\hfill \rm \null \hfill
 \hbox{\normalsize ADP-02-88/T527} \\
\vspace{-2mm}
 \hfill \hbox{\normalsize JLAB-THY-02-47} \\
\vspace{-1mm}
Light quark simulations with FLIC fermions}

\author{
J.~M.~Zanotti\address[CSSM]{Special Research Center for the
	Subatomic Structure of Matter, and		\\
	Department of Physics and Mathematical Physics,
	University of Adelaide, 5005, Australia}\thanks{Presented by
        J.~M.~Zanotti}, % at Lattice 2002},
D.~B.~Leinweber\addressmark[CSSM],
W.~Melnitchouk\address[JLab]{Jefferson Lab, 12000
	Jefferson Avenue, Newport News, VA 23606, U.S.A.},
A.~G.~Williams\addressmark[CSSM] and
J.~B.~Zhang\addressmark[CSSM]
}

\begin{document}

\begin{abstract}
  Hadron masses are calculated in quenched lattice QCD in order to
  probe the scaling behavior of a novel fat-link clover fermion action
  in which only the irrelevant operators of the fermion action are
  constructed using APE-smeared links.  Light quark masses corresponding
  to an $m_{\pi} / m_{\rho}$ ratio of 0.35 are considered to assess
  the exceptional configuration problem of clover-fermion actions.  This
  Fat-Link Irrelevant Clover (FLIC) fermion action provides scaling
  which is superior to mean-field improvement and offers advantages
  over nonperturbative improvement, including reduced exceptional
  configurations.
\end{abstract}

\maketitle

%\vspace{3mm}
%PACS number(s): 11.15.Ha, 12.38.Gc, 12.38.Aw

%\newpage

%%%%%%%%%%%%%%%%%%%%%%%%%%%%%%%%%%%%%%%%%%%%%%%%%%%%%%%%%%%%%%%%%%%%%%%%%%%
\section{INTRODUCTION}

The Sheikholeslami-Wohlert (clover) action
\cite{Sheikholeslami:1985ij} is given by

%introduces an additional irrelevant dimension-five operator to the
%standard Wilson \cite{Wilson} quark action in order to improve the
%scaling properties of the fermion action at finite $a$, 
%
\be
S_{\rm SW}
= S_{\rm W} - \frac{iC_{\rm SW} \kappa r}{2(u_{0})^4}\
	     \bar{\psi}(x)\sigma_{\mu\nu}F_{\mu\nu}\psi(x)\ ,
\label{clover}
\ee
where $S_{\rm W}$ is the standard Wilson action \cite{Wilson} and
$C_{\scriptstyle {\rm SW}}$ is the clover coefficient which can be
tuned to remove ${\cal O}(a)$ artifacts.
Nonperturbative (NP) ${\cal O}(a)$ improvement \cite{Luscher:1996sc}
tunes $C_{\scriptstyle {\rm SW}}$ to all powers in $g^2$ and displays
excellent scaling, as shown by Edwards {\it et al} \cite{Edwards:1998nh}.

However, the clover action is susceptible to the problem of
exceptional configurations as the quark mass becomes small.  In
practice, this prevents the use of coarse lattices ($\beta < 5.7 \sim
a > 0.18$~fm) \cite{Bardeen:1998gv,DeGrand:1998jq}.
Furthermore, the plaquette version of $F_{\mu\nu}$, which is commonly
used in Eq.~(\ref{clover}), has large ${\cal O}(a^2)$ errors, which
can lead to errors of the order of $10 \%$ in the topological charge
even on very smooth configurations \cite{Bonnet:2000dc}.

Initial studies \cite{FATJAMES} using a Fat-Link Irrelevant Clover
(FLIC) fermion action showed that, on a lattice with a spacing
corresponding to $a^2 \sigma \sim 0.08$, where $\sigma$ is the
string tension, it is possible to achieve superior scaling to that
using a mean-field improved clover action.
Furthermore, it is competitive with nonperturbatively improved clover
without having to fine tune the coefficients of action improvement.

In this paper we expand on previous results \cite{FATJAMES} by
examining the scaling of FLIC fermions at three different lattice
spacings. We also investigate the problem of exceptional
configurations by presenting preliminary results of simulations at
light quark masses corresponding to $m_{\pi}/m_{\rho} = 0.35$.

\vspace*{-0.1cm}
%%%%%%%%%%%%%%%%%%%%%%%%%%%%%%%%%%%%%%%%%%%%%%%%%%%%%%%%%%%%%%%%%%%%%%%%%%%
\section{GAUGE ACTION}
\label{simulations}

The simulations are performed using a tree-level ${\cal
  O}(a^2)$--Symanzik-improved \cite{Symanzik:1983dc} gauge action on
$12^3 \times 24$ and $16^3 \times 32$ lattices with lattice spacings
of 0.093, 0.122 and 0.165~fm determined from the string tension with
$\sqrt\sigma=440$~MeV.  A total of 200 configurations are used in the
scaling analysis at each lattice spacing and volume.  
In addition, for the light quark simulations, 94 configurations are
used on a $20^3 \times 40$ lattice with $a=0.134$~fm.
The error analysis is performed by a
third-order, single-elimination jackknife, with the $\chi^2$ per
degree of freedom ($N_{\rm DF}$) obtained via covariance matrix fits.

%%%%%%%%%%%%%%%%%%%%%%%%%%%%%%%%%%%%%%%%%%%%%%%%%%%%%%%%%%%%%%%%%%%%%%%%%%%
% \section{FAT-LINK IRRELEVANT CLOVER FERMION ACTION}
\section{FERMION ACTION}
\label{FLinks}

Fat links \cite{DeGrand:1998jq,DeGrand:1999gp} are created using
APE smearing \cite{ape} followed by a projection back to SU(3).
We repeat this procedure of smearing and projection $n$ times.
Fat links are created by setting $\alpha = 0.7$ and $n=4$. Further
details of FLIC fermion actions can be found in Ref.~\cite{FATJAMES}

As reported in Ref.~\cite{FATJAMES}, one finds that the plaquette
measure $u_0 \approx 1$ for the fat links, so that the mean-field
improved coefficient for $C_{\rm SW}$ is expected to be adequate.
Also, one can now use highly improved definitions of $F_{\mu\nu}$.
In particular, we employ an ${\cal O}(a^4)$ improved definition of
$F_{\mu\nu}$ \cite{sbilson} in which the standard clover-sum of four
$1 \times 1$ Wilson loops lying in the $\mu ,\nu$ plane is combined
with $2 \times 2$ and $3 \times 3$ Wilson loop clovers.
A fixed boundary condition is used for the fermions and
gauge-invariant gaussian smearing \cite{Gusken:qx} in the spatial
dimensions is applied at the source to increase the overlap of the
interpolating operators with the ground states.

\section{RESULTS}
\label{discussion}

\begin{figure}[t]
\begin{center}
{\includegraphics[height=\hsize,angle=90]{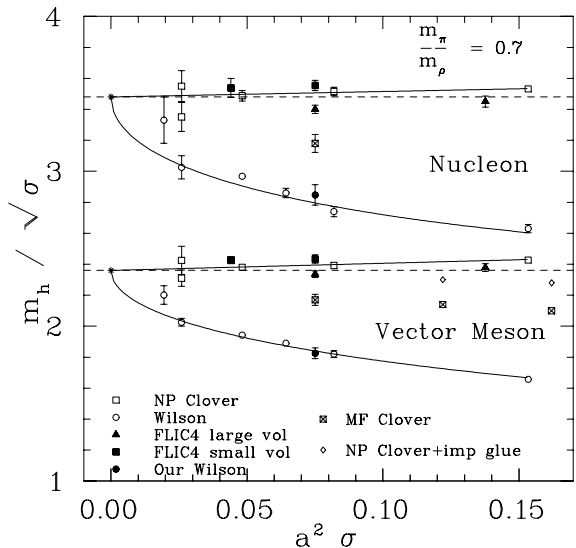}}
\vspace*{-1.0cm}
\caption{Nucleon and vector meson masses for the Wilson, NP-improved
  and FLIC actions obtained by interpolating the results to $m_\pi /
  m_\rho = 0.7$.  For the FLIC action (``FLIC4''), fat links are
  constructed with $n=4$ smearing sweeps at $\alpha = 0.7$.  Results
  from the present simulations are indicated by the solid symbols;
  those from earlier simulations by open or hatched symbols.
  \vspace*{-1.0cm} }
\label{scaling1}
\end{center}
\end{figure}

%Hadron masses are extracted from the Euclidean time dependence of the
%calculated two-point correlation functions in the standard way. 
%Effective masses are calculated as a function of time
%and various time-fitting intervals are tested with a covariance matrix
%to obtain $\chi^2 / N_{\rm DF}$.
%
The scaling behavior of the various actions is illustrated in
Fig.~\ref{scaling1}.  The present results for the Wilson action agree
with those of Ref.~\cite{Edwards:1998nh}.
The FLIC action performs systematically better than the mean-field
improved clover, indicating that only 4 sweeps of smearing, combined
with our ${\cal O}(a^4)$ improved $F_{\mu\nu}$, provides excellent
results, without the fine tuning of $C_{\rm SW}$ in the NP improvement
program.
The FLIC results also compete well with those obtained with the
NP-improved clover fermion action.

%By examining the results in Fig.~\ref{scaling1} as a function of $a^2
%\sigma$, we immediately see that there is only a small change in the
%results as we vary the lattice spacing. This would seem to indicate
%that by smearing the irrelevant operators in our fermion action, we
%have removed most of the ${\cal O}(a)$ and ${\cal O}(a^2)$ errors.
%
The two different volumes used at $a^2 \sigma \sim 0.08$
suggest a small finite volume effect, which increases
the mass for the smaller volume. 
Examination of points on the small and large volumes separately
indicates scaling consistent with errors of ${\cal O}(a^2)$.
This contrasts with mean-field improved results, where an
extrapolation of the vector meson mass linear in $a^2 \sigma$
cannot accommodate the continuum limit estimate.
These results indicate that FLIC fermions provide a
new form of nonperturbative ${\cal O}(a)$ improvement.
%
%This would seem to further enhance the previous statement
%because if a larger volume lattice were used at $a^2 \sigma \sim
%0.045$, we would expect the point to drop down slightly, consistent
%with results for the larger volumes performed at the coarser lattice
%spacings. Of course, to rigorously test this statement one should do
%this calculation at different lattice spacings holding the volume
%constant and then repeat the process for different volumes. However,
%this would be extremely time consuming and we probably wouldn't
%learn much more than we have with our present simulations.

%Finally, in order to search for exceptional configurations by pushing
%the bare quark mass down, we would like our preferred action to be
%efficient when inverting the fermion matrix. 
Previous work \cite{FATJAMES,WASEEM} has shown that the FLIC fermion
action has extremely impressive convergence rates for matrix
inversion, which provides great promise for performing cost effective
simulations at quark masses closer to the physical values.  Problems
with exceptional configurations have prevented such simulations in the
past.

\begin{figure}[t]
\begin{center}
{\includegraphics[height=\hsize,angle=90]{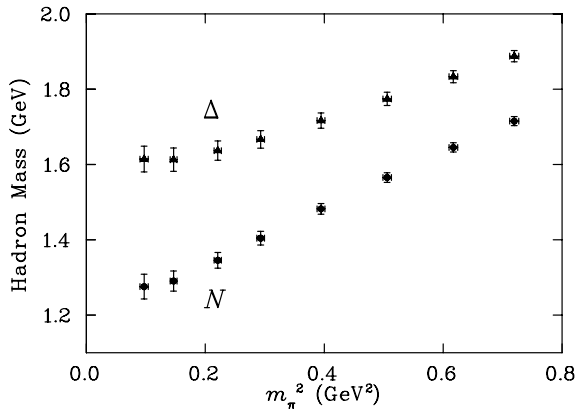}}
\vspace*{-1.0cm}
\caption{Nucleon and $\Delta$ masses for the FLIC action on
  a $20^3 \times 40$ lattice with $a=0.134$~fm. 94 configurations were
  used and the fat links are constructed with $n=6$ smearing sweeps at
  $\alpha = 0.7$.  \vspace*{-1.0cm} }
\label{LQM}
\end{center}
\end{figure}

The ease with which one can invert the fermion matrix using FLIC
fermions leads us to attempt simulations down to small quark masses
corresponding to $m_{\pi} / m_{\rho} = $ 0.35. Previous attempts with
Wilson-type fermion actions have only succeeded in getting down to
$m_{\pi} / m_{\rho} = 0.47$ \cite{TMass}. The simulations are on a
$20^3 \times 40$ lattice with a physical length of 2.7~fm. We have
used an initial set of 94 configurations, using $n=6$ sweeps of
smearing, and preliminary results indicate exceptional
configurations at the 1\% level \cite{LQMJAMES}.
Figure~\ref{LQM} shows the $N$ and $\Delta$ masses as a function of
$m_{\pi}^2$ for all eight masses considered.
It is interesting to note an upward curvature in the $\Delta$ mass,
increasing the $N-\Delta$ mass spitting for decreasing quark mass.
This behavior has been predicted by Young {\em et al} \cite{ROSS}.
%
%The first two of these lighter quark
%masses show no sign of exceptional configurations, however at the
%lightest mass tested, the pion correlator exhibited divergences on
%$1\%$ of the configurations. These exceptional configurations were
%omitted when producing Fig.~\ref{LQM}.

%%%%%%%%%%%%%%%%%%%%%%%%%%%%%%%%%%%%%%%%%%%%%%%%%%%%%%%%%%%%%%%%%%%%%%%%%%%
% \section{Conclusions and Future Work}
\section{CONCLUSIONS}
\label{conclusion}

We have calculated hadron masses to test the scaling of a novel
Fat-Link Irrelevant Clover fermion action, in which only the
irrelevant, higher-dimension operators involve smeared links.
One of the main conclusions of this work is that the use of fat links
in the irrelevant operators provides a new form of nonperturbative
${\cal O}(a)$ improvement.
This technique competes well with ${\cal O}(a)$ nonperturbative
improvement on mean field-improved gluon configurations, with the
advantage of a reduced exceptional configuration problem.

Quenched simulations at quark masses down to $m_{\pi}/m_{\rho}=0.35$
have been successfully performed on a $20^3 \times 40$ lattice with
a lattice spacing of 0.134~fm.
Simulations at such light quark masses hold promise for revealing
non-analytic behaviour of quenched chiral perturbation theory.

%%%%%%%%%%%%%%%%%%%%%%%%%%%%%%%%%%%%%%%%%%%%%%%%%%%%%%%%%%%%%%%%%%%%%%%%%%%
% \acknowledgements
\vspace*{0.3cm}

We thank the Australian National Computing Facility for
Lattice Gauge Theory, and the Australian Partnership for Advanced
Computing (APAC) for supercomputer time.
This work was supported by the Australian Research Council.
W.M. was supported by the U.S. Department of Energy contract
\mbox{DE-AC05-84ER40150}, under which the Southeastern Universities
Research Association (SURA) operates the Thomas Jefferson National
Accelerator Facility (Jefferson Lab).

%%%%%%%%%%%%%%%%%%%%%%%%%%%%%%%%%%%%%%%%%%%%%%%%%%%%%%%%%%%%%%%%%%%%%%%%%%%

\end{document}